\documentclass[journal=jacsat,manuscript=article]{achemso}
\usepackage{chemformula} 
\usepackage[normalem]{ulem}
\usepackage[T1]{fontenc}

\author{Patrice Theul\'e}
\affiliation{Aix Marseille Univ, CNRS, CNES, LAM, Marseille, France}
\email{patrice.theule@univ-amu.fr}
\author{Christian Endres}
\altaffiliation{Current address: MPE Garching}
\author{Marius Hermanns}
\affiliation{I. Physikalisches Institut, Universit\"at zu K\"oln, Zulpicher Str. 77, D-50937 K\"oln, Germany}
\author{Jean-Baptiste Bossa}
\affiliation{Raymond and Beverly Sackler Laboratory for Astrophysics, Leiden Observatory, Leiden University, P.O. Box 9513, NL 2300 RA Leiden, The Netherlands}
\author{Alexey Potapov}
\affiliation{I. Physikalisches Institut, Universit\"at zu K\"oln, Zulpicher Str. 77, D-50937 K\"oln, Germany}
\altaffiliation{Current address: Laboratory Astrophysics Group of the Max Planck Institute for Astronomy 
  at the Friedrich Schiller University Jena, Institute of Solid State Physics, Helmholtzweg 3, 07743 Jena, Germany}
\email{alexey.potapov@uni-jena.de}

\title[] {High-resolution gas phase spectroscopy of molecules desorbed from an ice surface:\\ a proof-of-principle study}
\abbreviations{MW, MMW, CP, FT, FID, TPD, ML}

\begin{document}

\begin{abstract}
High-resolution gas phase spectroscopy techniques in the microwave, millimeter-wave and terahertz spectral ranges can be used to study complex organic molecules desorbed from interstellar ice analogues surface with a high sensitivity. High-resolution gas phase spectroscopy gives unambiguous information about the molecular composition, the molecular structure, and transition frequencies needed for their detection by radio telescopes in various interstellar and circumstellar environments. The results will be useful not only for interpreting astronomical spectra and understanding astrophysical processes, but also for more general studies of gas-surface chemistry. This paper presents a new experimental approach based on a combination of a chirped-pulse Fourier transform microwave spectrometer detection and a low temperature surface desorption experiment. The experimental set-up is benchmarked on the desorption of ammonia ice detected by high-resolution gas phase microwave spectroscopy.
\end{abstract}

\section{Introduction}

Ices in dense molecular clouds, in protostellar envelopes and in protoplanetary disks are at the origin of the complex organic molecules (COMs) that cannot be created via gas phase reactions. These COMs originate from a few simple molecules observed in interstellar ices, such as H$_2$O, NH$_3$, CH$_4$, CO, CO$_2$, and they give us a great insight in the astrochemical processes at work in the different regions they are observed in. Of particular interest is, of course, their formation pathway in star forming regions \cite{CaselliAAR12,CeccarelliApJ17}.
Vibrational spectral analysis of observed interstellar ice IR spectra  has enabled the unambiguous identification of at least six molecules (and some of their isotopologues) as well as tentative detection of several others \cite{BoogertARAA15}. Compared to pure gas phase spectra, ice bands are broad, with significant spectral overlapping, which make their assignment and subsequent molecule identification difficult and limited. In addition, absorption of IR radiation by the terrestrial atmosphere makes ground-based observatories not suitable for ice spectroscopy and satellite or airborne telescopes must be used. As a result far more molecules, about 200, have been detected in the gas phase by radio astronomy \cite{MullerAA01} than have been found in the solid state.

In the laboratory, “solid” COMs have been typically synthesized in molecular ices using the main triggers of grain surface chemistry at work in interstellar and circumstellar environments, such as thermal processing, UV and X-ray irradiation, atom, ion, proton, and electron bombardment (for review see \cite{TheuleASR13,LinnartzIRPC15,ObergCR16}). Recently, an alternative route of COMs formation - implying dust grain surface chemistry – has been experimentally demonstrated \cite{PotapovApJ17}. The two main laboratory methods for studying molecules produced in surface or bulk reactions are solid state Fourier transform infrared (FTIR) spectroscopy and gas phase mass spectrometry (MS). FTIR spectrometers combine many advantages such as multiplex recording, broadband coverage, and easy wavelength calibration. Mass spectroscopy has a very high sensitivity, with typical limit of detection 10$^{12}$ molecules m$^{-2}$ s$^{-1}$ and capable of reading partial pressures down to $\sim$ 10$^{-13}$ mbar \citep{FraserRSI02}. These methods have been used in a huge amount of studies on the modelling of astrophysically relevant surface and solid-state processes (see the review papers mentioned above and references therein).

However, these methods have limitations. At smaller wavenumbers, in the terahertz frequency range, which is the range of many ground-based observatories (e.g. Nobeyama, LMT, IRAM 30m, ALMA) the signal to noise of the FTIR spectrometers is limited due to the low power density of the thermal radiation source. In addition, IR spectral bands are weak and broad, often overlapping with more intense bands corresponding to reactant species. The main problem of MS is the low mass resolution of most of the mass spectrometers typically used. In addition, in standard MS experiments it is not possible to distinguish between molecules of the same mass, or isomers of the same molecule. This has been recently overcome by fragment-free tunable single photon vacuum ultraviolet photoionization in combination with reflectron time-of-flight mass spectrometry \cite{AbplanaplCPL16}; however, this method is not widely available. More fundamentally, the MS techniques are limited to the solar system, and most of the ISM studies rely on photometric or spectroscopic observations. Solar system missions, such as Cassini-Huygens \cite{CoustenisRAA09} and Rosetta \cite{KrugerPSS15}, used mass spectrometry but so far the identification of complex molecules has been limited \cite{AltweggMNRAS17}. However, an impressive result was a first unambiguous detection of glycine in extraterrestrial medium, found in the coma of comet 67P/Churyumov-Gerasimenko \cite{AltweggSA16}.

The problems of the FTIR and MS methods mentioned above do not exist in high-resolution spectroscopy in the microwave (MW), millimeter-wave (MMW), and terahertz (THz) spectral regions. High-resolution laboratory spectra can be directly compared to spectra observed by ground-based radio observatories and give unambiguous information on the molecular composition and structure of molecules. In addition, some astronomically relevant COMs, known to be produced in solid, are not stable at room temperature in the gas phase. A possibility to obtain their spectroscopic signatures is to synthesize these molecules in interstellar ice analogues and detect them above the surface of ice immediately after their release into the gas phase. There is, to the best of our knowledge, only one very recent example of gas phase high-resolution spectroscopy used in studies of interstellar ice analogues \citep{YocumJCPA19}. The setup is based on a broadband high-resolution THz spectrometer that allows observations of spectra of desorbed molecules, H$_2$O and CH$_3$OH, directly above the ice surface.

In this paper, we discuss an innovative approach for the laboratory study of ``solid'' molecules - a combination of a low temperature solid state experiment and gas phase high-resolution spectroscopy. This combination should eliminate the existing limitations of laboratory measurements of molecules, especially COMs, generated in cosmic ice analogues. The approach is aimed at measurements of high-resolution gas phase spectra of important astronomically relevant molecules known to be formed in the solid state for which there are no spectral reference data available - for example, NH$_2$COOH \citep{BarattaAA02}, NH$_2$CH$_2$OH \cite{BertinPCCP09}, HOCH$_2$OH \cite{BossaAA09}, NH$_4$NH$_2$COO$^-$ \cite{BossaAA08,PotapovApJ19}. We present here proof-of-principle measurements and calculations based on the thermal desorption of ammonia ice detected by a chirped-pulse Fourier transform microwave spectrometer.  

\section{Experimental part}

\begin{figure}
\resizebox{\hsize}{!}{\includegraphics[scale=0.4]{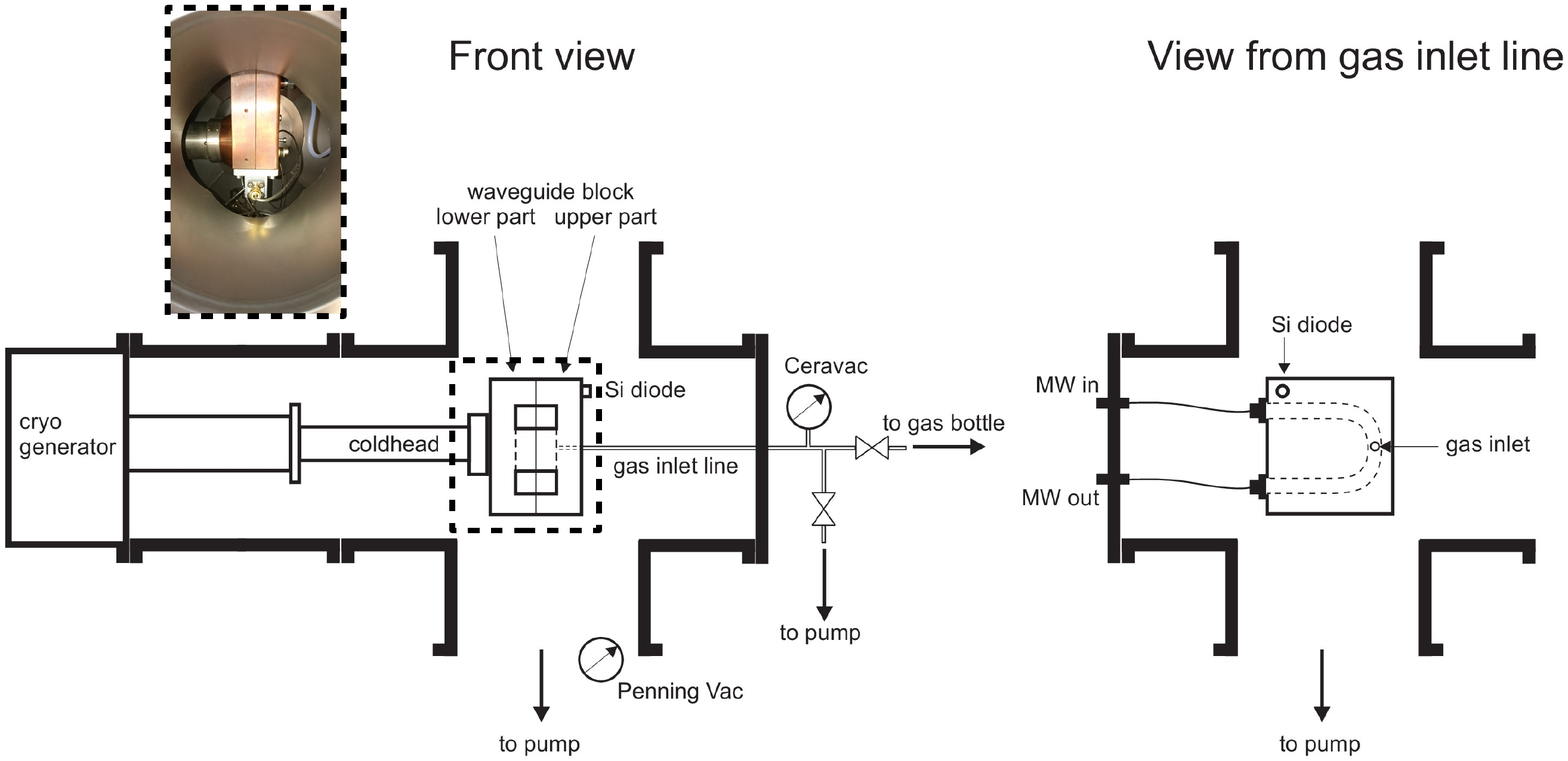}}
\caption{Schematic of the experimental set-up main chamber. A U-shaped waveguide is cooled down by cryogenerator. Gas is inserted in the waveguide through a capillary tube. The waveguide temperature is measured by a Si-diode. The vacuum pressure is measured both in the chamber and in the gas inlet line. The MW radiation is injected at one end of the waveguide and the free induction decay (FID) of the molecules studied is detected at the other end. Inset: photograph of the cooled waveguide inside the main chamber (dashed line zone).}
\label{chamber}
\end{figure}

The set-up developed in the I. Physical Institute of the University of Cologne is a combination of a chirped-pulse Fourier transform (CP-FTMW) spectrometer \cite{HermannsJMS19} and a U-shaped waveguide as a molecular cell mounted in a high-vacuum chamber (10$^{-6}$ mbar at room temperature and 10$^{-6}$ mbar at 20 K), where it can be cooled to cryogenic temperatures (down to 20 K). The ice sample is deposited onto the waveguide inner walls and molecules desorbed are detected inside the cold waveguide by the CP-FTMW spectrometer.
The U-shaped waveguide is mounted to the cold head, as shown in Fig. \ref{chamber}. It is a Ku band waveguide (WR-62), machined in split block technique from two copper pieces, which are tightly screwed. Each half has a 42.0 mm length and width and a 7.9 mm height. The waveguide curvature has a 52.0 mm outer diameter and a 36.2 mm inner diameter. It has a volume of 19.1 cm$^{3}$ and an inner surface of 72.6 cm$^2$. The inner surfaces are mechanically polished. A DT-670 silicon diode (Lakeshore Inc.) measures the temperature with a 0.5 K accuracy on the outer part of the waveguide as schematized in Fig. \ref{chamber}.

The CP-FTMW technique has been developed by Brooks Pate’s group \cite{BrownRSI08}, as an improvement of the Fourier transform microwave technique \cite{BalleRSI81}, and it has been regularly improved since then \cite{ParkJCP11, JahnJMS12,SteberJMS12, OldhamJCP14, ProzumentPCCP14} working now not only in the MW but also in the MMW and THz spectral regions.
The spectrometer in Cologne operates in a frequency range from 2 to 50 GHz \cite{HermannsJMS19} obtained by mixing a variable frequency of a microwave synthesizer with a chirp of a linear frequency sweep of 1 GHz generated by an arbitrary waveform generator, following recent developments \cite{JahnJMS12}, and was used in the present study in the frequency range around 24 GHz.
The microwave pulse is broadcasted into the waveguide. Molecules under study, which transitions are falling into the chosen frequency range, are polarized by the chirped pulse and undergo a broadband free induction decay (FID), which is received by an antenna, amplified, down-converted and digitized by a fast broad bandwidth oscilloscope. Phase-synchronized repetition is used for phase coherent averaging of the signal in the time domain. An averaged spectrum is then fast Fourier transformed into a frequency domain spectrum. Typical parameters used in the experiments presented here were a 700 nanoseconds chirped pulse width, a 1 GHz frequency span, and 1000 averaged scans, which represents about 2 second acquisition time.

We benchmarked our experimental set-up recording the inversion spectrum of desorbing ammonia ice. The NH$_3$ gas (purity 5.5) was first prepared in a primary pumped vacuum injection line at a pressure of a few mbar measured by a capacitance vacuum gauge. The gas was inserted through a 1.0 mm diameter inlet hole in the waveguide connected to the gas inlet line by a teflon tube for thermal decoupling. The gas was injected in the cold waveguide at a pressure of 10$^{-6}$ mbar for few tens of seconds (which also deposits water and other gases as pollutants), and subsequently frozen on the inner surface of the waveguide. The amount of material frozen in the waveguide can be determined by the injection time and the dosing pressure; in practice by following the pressure in the dosing line as a function of time and assuming a sticking coefficient of one. Supposing that the injected molecules are uniformly deposited over the 72.6 cm$^2$ inner surface of the waveguide, the number of molecules was converted into surface coverage, with an assumption of 10$^{15}$ molecules per cm$^2$ per monolayer (ML). Vacuum gauges give only a rough estimate, the thickness was also calibrated by integrating the microwave line intensities during the molecular desorption.

After the deposition, the waveguide temperature was linearly ramped to room temperature in a temperature programmed desorption (TPD) experiment with the ramp rate of 0.4 K min$^{-1}$. The spectrum of the desorbing species was acquired during the temperature ramp using the CP-FTMW spectrometer. The intensity of the molecular spectrum was a measure for the amount of gas present in the waveguide. The ice sublimation was therefore monitored by the gas specific spectral features and by the pressure gauge on the setup.

\section{Results}

\begin{figure}
\includegraphics[scale=0.4]{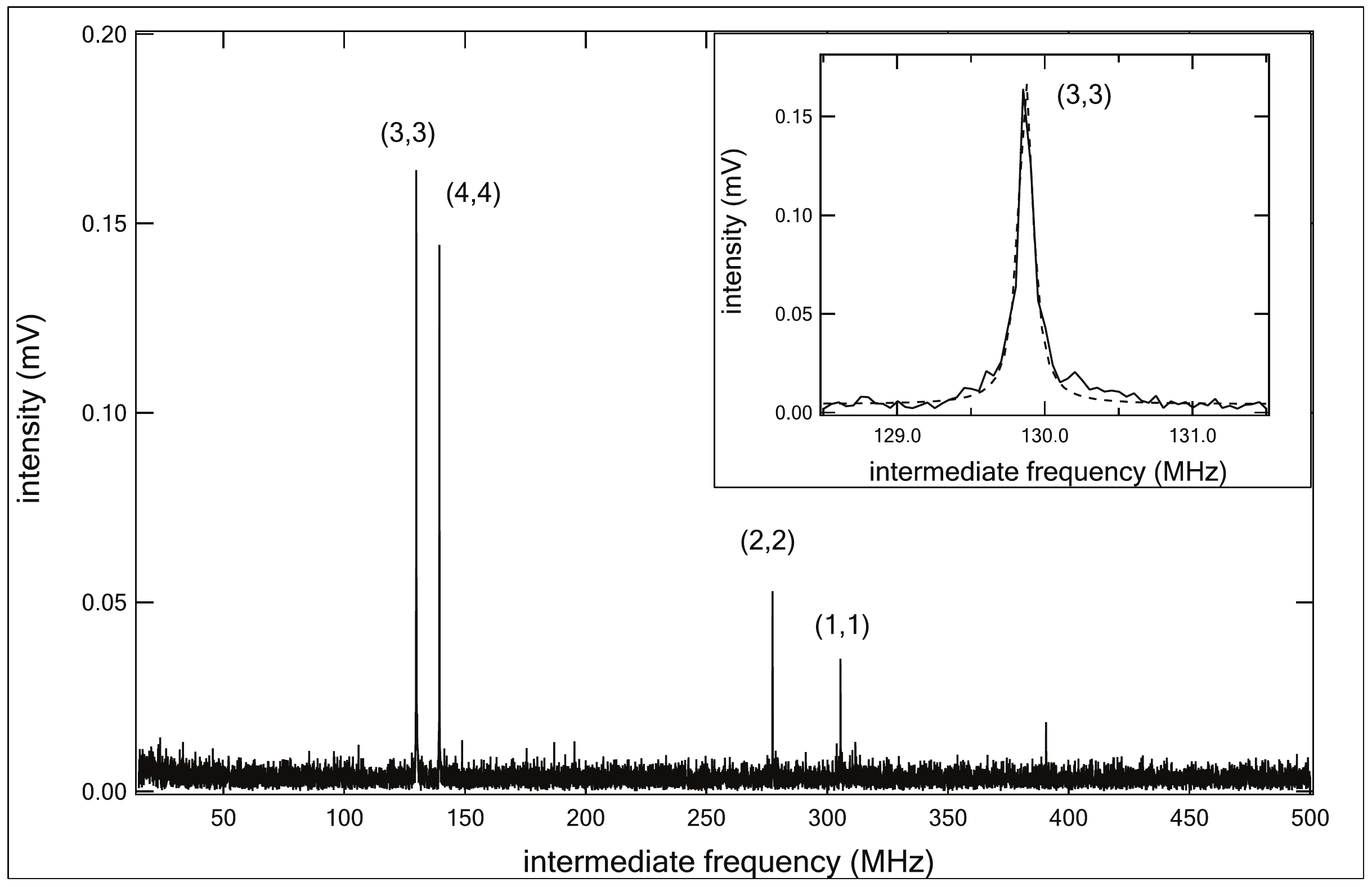}
\caption{Inversion spectrum of NH$_3$ between 23.5 GHz and 24.5 GHz acquired during the desorption of 1.4 monolayer of NH$_3$ at 110 K. The synthesizer frequency was set to 24 GHz; both the lower and upper side bands are displayed. The (J,K)= (1,1), (2,2), (3,3) and (4,4) inversion transitions are recorded during a 1 GHz frequency sweep. Inset: fitting the (3,3) inversion transition with a Lorentzian function gives a 120 kHz linewidth.}
\label{spectrum}
\end{figure}

The CP-FTMW spectrometer was used, for the first time, to obtain TPD data on ice molecules resembling those in the interstellar medium that were released to the gas phase. The inversion spectrum of thermally desorbed ammonia molecules deposited on the walls of the waveguide at 20 K was monitored as the waveguide was warmed up at a constant rate. The inversion spectrum NH$_3$ is caused by the tunneling of N atom through the plane of three H atoms resulting in a splitting of the energy levels. Thus, each rotational state, described by the rotational quantum numbers J and K, gives rise to a transition in the inversion spectrum. A frequency span of 1 GHz allowed recording four inversion transitions of ammonia in a single data acquisition: (J,K) = (1,1) at 23694.4955MHz, (2,2) at 23722.6333MHz, (3,3) at 23870.1292MHz, and (4,4) at 24139.4163MHz. 
Figure~\ref{spectrum} shows the inversion spectrum of NH$_3$ between 23500 MHz and 24500 MHz acquired during the desorption of 1.4 monolayer of NH$_3$ at 110 K. The synthesizer frequency was set to 24 GHz; both the lower and upper side bands are displayed.

\begin{figure}
\includegraphics[scale=0.5]{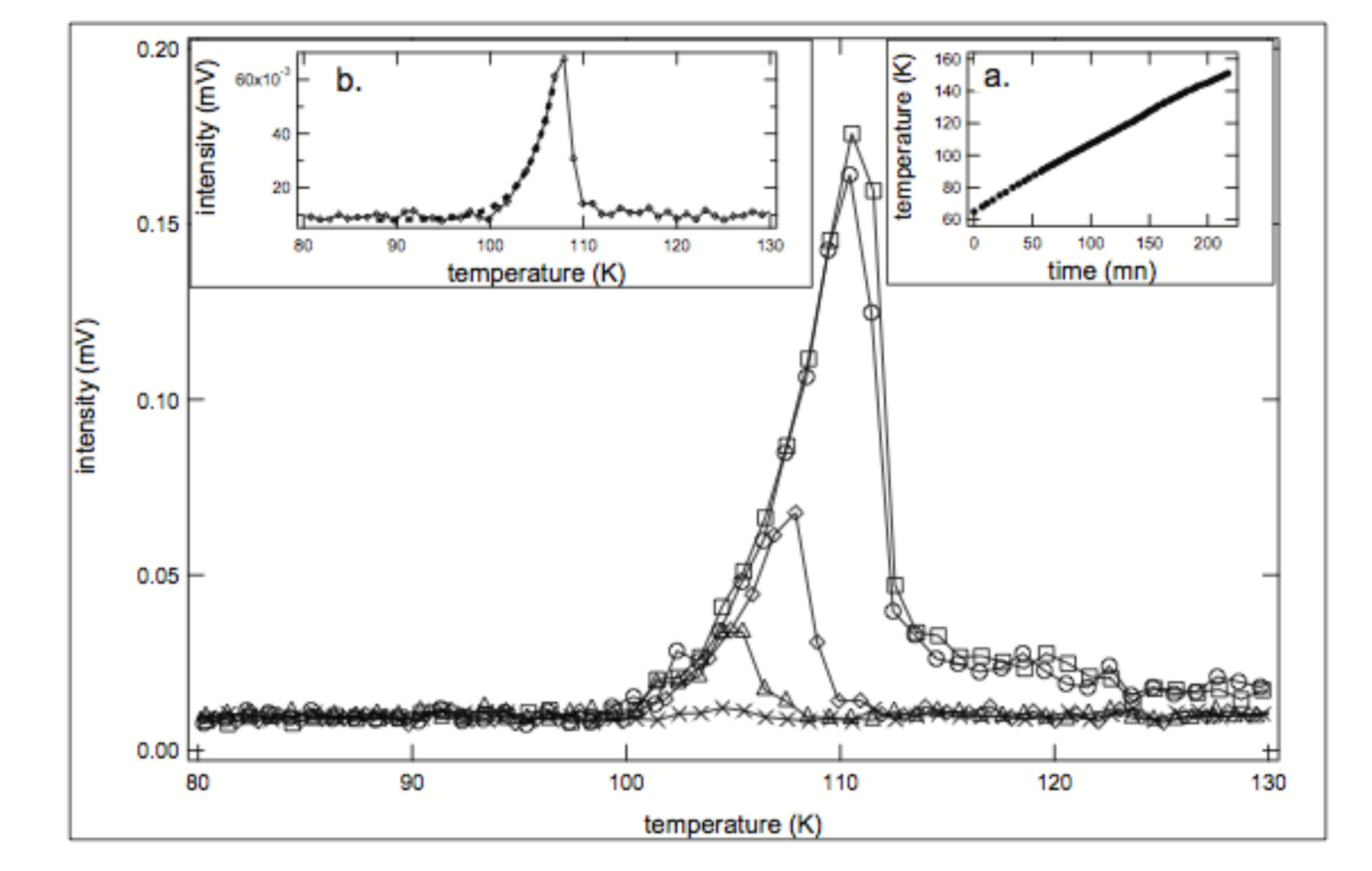}
\caption{TPD curves for NH$_3$ ices with 1.5 ML (open squares), 1.4 ML (open circles), 0.8 ML (open diamonds), 0.4 ML (open triangles), and 0.1 ML (crosses) thickness. The FID signal of the (3,3) transition was recorded as a function of temperature, which was ramped linearly with a 0.4 K/min ramp rate (inset a). The NH$_3$ desorption curve for 0.8 ML is fitted with the Polanyi-Wigner equation (dashed lines, inset b).}
\label{TPD}
\end{figure}

The TPD curves obtained by taking the FID signal of the (3,3) transition as a function of temperature are presented in Figure \ref{TPD}.

The onset of desorption in our TPD curves is shifted compared to the literature mass spectroscopy data \cite{CollingsMNRAS04}, where the TPD curve of NH$_3$ peaks at about 95 K and no signal is observable above 105 K. However, previous FTIR measurements showed a slow (on an hour timescale) desorption of NH$_3$ ice at 100, 103, and 105 K \cite{SandfordApJ93}. Thus, our results are more in agreement with \cite{SandfordApJ93}.
One can see zeroth-order kinetics of desorption in the coverage range of 1.5 – 0.4 ML and first-order kinetics of desorption in the range of 0.4 – 0.1 ML. Typically, the first relates to the multilayer desorption and second to the desorption of individual molecules or monolayer desorption. Thus, one could expect first-order desorption kinetics in the whole coverage range presented here. A shift of the transition point to thinner ice samples can be explained by the fact that in our experiments molecules desorb not from an opened surface when they are immediately pumped out, but are confined in the waveguide with a limited pumping capacity \citep{FraserMNRAS01,AcharyyaAA07}. The gas flow rate from the waveguide has been measured by initially filling the waveguide with NH$_3$ at room temperature and recording the pressure decay rate. The leak rate of the overall waveguide system to the main chamber resulted in a pumping speed of 5$\times$10$^3$ L/s. This may explain also a slight shift of the desorption temperature. This temperature shift should be decreased by a more efficient pumping, either by using a more powerful main chamber pump or by increasing the waveguide opening to the main chamber. Another explanation for the temperature shift is the presence of water pollution in the NH$_3$ ice. A delayed desorption due to the wetting on a water surface has already been observed for H$_2$CO \citep{NobleAA12}.

We fit\sout{ted} the microwave TPD curves (except 0.4 ML) with the zeroth-order Polanyi-Wigner equation \cite{RedheadV62}:
\begin{equation}
\frac{dN}{dT}=-\frac{1}{\beta}\nu_0\exp(-\frac{Edes}{RT})
\end{equation}

where N is number of molecules, T – temperature, $\beta$ – ramp rate, $\nu_0$ – pre-exponential factor, and $E_{des}$ – desorption energy. Fixing the pre-exponential factor to 10$^{12}$ s$^{-1}$, we obtained $E_{des}$ = 31.6 $\pm$ 0.1 kJ mol$^{-1}$, which is in a reasonable agreement with the value of 25.6 kJ mol$^{-1}$ obtained by FTIR spectroscopy \cite{SandfordApJ93} taking into account different experimental procedures of two experiments. 

\begin{figure}
\resizebox{\hsize}{!}{\includegraphics[width=16cm, height=9cm]{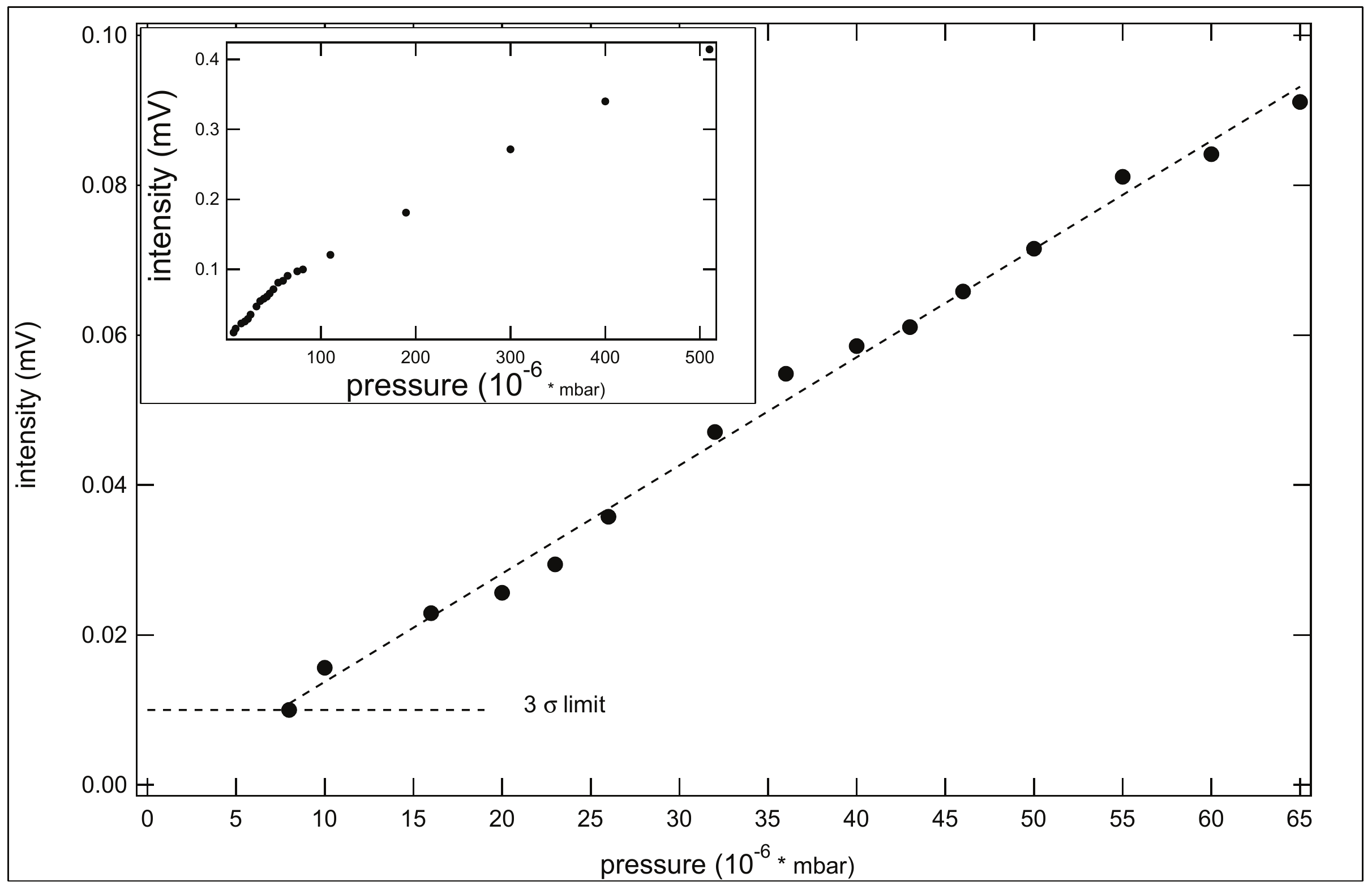}}
\caption{Intensity of the (3,3) transition as a function of the NH$_3$ pressure in the waveguide. The microwave signal is linear to the quantity of molecules. The slope change observed near 10$^{-4}$ mbar corresponds to switching between the Pirani gauge and the cold cathode Penning gauge. The 3$\sigma$ limit of the signal sets the sensitivity limit of the set-up to 6 10$^{-6}$ mbar or 5 picomoles.}
\label{sensitivity}
\end{figure}

To quantify the sensitivity of our set-up, we filled both the chamber and the waveguide with a defined pressure of NH$_3$. A fast pressure equilibrium between the waveguide inner volume and the vacuum chamber was obtained by removing the teflon tube between the gas inlet line and the waveguide section. The pressure was read from the pressure gauge of the main vacuum chamber. We monitored the intensity of the (3,3) line, which is the most intense line at room temperature in the given frequency range. 
The NH$_3$ pressure was decreased until the (3,3) line intensity reached the 3$\sigma$ limit as displayed in Figure~\ref{sensitivity}. This was obtained for c.a. 6$\times$10$^{-6}$ mbar. Within the ideal gas approximation, 6$\times$10$^{-6}$ mbar corresponds to 3$\times$10$^{12}$ molecules in 19.1 cm$^3$ of the waveguide or about 5 picomoles.

From Figure~\ref{sensitivity} we derived a conversion factor of 1.4$\times$10$^{-3}$ mV/(1$\times$10$^{-6}$ mbar)=1.4 V/mbar, which allowed us to scale Figure~\ref{TPD} into mbar, similarly to TPD curves derived from quadrupole mass spectrometry. Integrating the TPD curve over time, taking into account the 0.4 K/min ramp rate, we derived a total number of desorbed molecules and converted it into monolayer units on the basis of the assumed surface density of 10$^{15}$ molecules cm$^{-2}$ per monolayer and the inner waveguide surface of 72.6 cm$^{-2}$. Doing that, we found the initial number of deposited molecules in agreement with the number estimated from the manometric technique as described in the experimental section, which validates our procedures.

\section{Discussion}

With a picomol sensitivity, our set-up, compared with mass spectrometry techniques \citep{FraserRSI02}, has the advantage of providing broadband, 1 GHz, molecular spectra of desorbing volatiles, additionally selected from their desorption energies. The high sensitivity comes both from the use of a waveguide, which confines the electric field, and the use of the CP-FTMW technique, which enables a multiple integration of spectra due to the possibility of a phase reproducible excitation. 
With a 72.6 cm$^2$ inner surface of the waveguide, considering 10$^{15}$ molecules cm$^{-2}$ per monolayer, one ice monolayer coating will provide 100 nanomoles inside the waveguide, which is five orders of magnitude more than the detection limit. Scaling to a smaller dipole moment of a molecule (e.g. 0.1 D compared to 1.4718 D for NH$_3$) and taking into account that solid-state reactions usually have a low yield, on the order of a percent, we are still two orders of magnitude above the detection limit. Moreover, the sensitivity can be further increased by increasing the integration time and the power of the microwave excitation. Thus, our set-up should be sensitive enough to study rotational spectra of complex organic molecules formed in the solid state and released into the gas phase. 
Due to the spectral broadness of the microwave pulse, it should be possible to obtain reasonable values for relative line intensities recorded in a single data acquisition.
However, first, the intensity of the microwave power should be calibrated. To do this, a number of effects, such as chirped pulse duration \cite{AbeysekeraJCP14}, edge effects \cite{ParkJMS15}, sideband gains, and frequency- and temperature-dependent losses in the waveguide, have to be taken into account. This task was out of the borders of the study presented here.

In the experimental set-up we developed the use of a waveguide greatly increases the detection sensitivity, which is critical for the analysis of COMs. However, it has a number of limitations, most significantly the number of different triggers of surface reactions leading to the formation of COMs. The main triggers of grain surface chemistry in interstellar and circumstellar environments are  thermal processing, UV irradiation, cosmic ray bombardment, and atom bombardment. Only thermal processing can be studied in our set-up with the waveguide in its current configuration. Thermal processing is very important particularly in protostellar envelopes and protoplanetary disks leading to the formation of many COMs \citep{TheuleASR13}. However, our set-up could be improved to enable VUV irradiation or atom/ion bombardment of the ice surface. For example by adding a UV window flange to the waveguide, a UV beam can illuminate the whole waveguide surface to (i.) photo-induce solid-state reactions, (ii.) photo-desorb solid-state species, (iii.) photo-produce gas-phase radical that may physisorb on the cold waveguide surface. Another solution is to make one part of the waveguide movable and fix it on a translator. In such a configuration, ice will be deposited onto one half of the waveguide and irradiated/bombarded. After that, the waveguide will be closed to the second half using a translator without breaking the vacuum in the chamber and TPD - chirped pulse experiments will be performed. In all these cases, rotational spectroscopy characterization of the desorbed molecules has lots of advantages with respect to mass spectrometry.  

\section{Conclusion}

We present a new laboratory experimental approach aimed at high-resolution spectroscopic measurements of molecules produced in interstellar ice analogues and released into the gas phase. Temperature programmed desorption of NH$_3$ ice deposited on the walls of a waveguide at 20 K was followed by recording the microwave inversion spectrum of ammonia using a chirped-pulse Fourier transform spectrometer. The results presented show a clear perspective of the approach for the gas phase detection of complex organic molecules produced in interstellar and circumstellar ice analogues.

\begin{acknowledgement}

Work of the authors in Cologne was supported by professor Stephan Schlemmer and the Deutsche Forschungsgemeinschaft (grant SFB 956, B3 and B4).
PT thanks the Deutsche Forschungsgemeinschaft for fundings his stays, and Prof. Dr. Schlemmer for hosting him, and Prof. J.S. Muenter for discussions.
J.-B.B. is grateful for support from the Marie Sklodowska Curie actions and the Intra-European Fellowship (FP7-PEOPLE-2011-IEF-299258). 

\end{acknowledgement}

\bibliography{biblio_papericemw_ACS}

\providecommand{\latin}[1]{#1}
\makeatletter
\providecommand{\doi}
  {\begingroup\let\do\@makeother\dospecials
  \catcode`\{=1 \catcode`\}=2 \doi@aux}
\providecommand{\doi@aux}[1]{\endgroup\texttt{#1}}
\makeatother
\providecommand*\mcitethebibliography{\thebibliography}
\csname @ifundefined\endcsname{endmcitethebibliography}
  {\let\endmcitethebibliography\endthebibliography}{}
\begin{mcitethebibliography}{37}
\providecommand*\natexlab[1]{#1}
\providecommand*\mciteSetBstSublistMode[1]{}
\providecommand*\mciteSetBstMaxWidthForm[2]{}
\providecommand*\mciteBstWouldAddEndPuncttrue
  {\def\EndOfBibitem{\unskip.}}
\providecommand*\mciteBstWouldAddEndPunctfalse
  {\let\EndOfBibitem\relax}
\providecommand*\mciteSetBstMidEndSepPunct[3]{}
\providecommand*\mciteSetBstSublistLabelBeginEnd[3]{}
\providecommand*\EndOfBibitem{}
\mciteSetBstSublistMode{f}
\mciteSetBstMaxWidthForm{subitem}{(\alph{mcitesubitemcount})}
\mciteSetBstSublistLabelBeginEnd
  {\mcitemaxwidthsubitemform\space}
  {\relax}
  {\relax}

\bibitem[{Caselli} and {Ceccarelli}(2012){Caselli}, and
  {Ceccarelli}]{CaselliAAR12}
{Caselli},~P.; {Ceccarelli},~C. {Our astrochemical heritage}. \emph{The
  Astronomy and Astrophysics Review} \textbf{2012}, \emph{20}, 56\relax
\mciteBstWouldAddEndPuncttrue
\mciteSetBstMidEndSepPunct{\mcitedefaultmidpunct}
{\mcitedefaultendpunct}{\mcitedefaultseppunct}\relax
\EndOfBibitem
\bibitem[{Ceccarelli} \latin{et~al.}(2017){Ceccarelli}, {Caselli}, {Fontani},
  {Neri}, {L{\'o}pez-Sepulcre}, {Codella}, {Feng}, {Jim{\'e}nez-Serra},
  {Lefloch}, and {Pineda}]{CeccarelliApJ17}
{Ceccarelli},~C.; {Caselli},~P.; {Fontani},~F.; {Neri},~R.;
  {L{\'o}pez-Sepulcre},~A.; {Codella},~C.; {Feng},~S.; {Jim{\'e}nez-Serra},~I.;
  {Lefloch},~B.; {Pineda},~J.~E. {Seeds Of Life In Space (SOLIS): The Organic
  Composition Diversity at 300-1000 au Scale in Solar-type Star-forming
  Regions}. \emph{The Astrophysical Journal} \textbf{2017}, \emph{850},
  176\relax
\mciteBstWouldAddEndPuncttrue
\mciteSetBstMidEndSepPunct{\mcitedefaultmidpunct}
{\mcitedefaultendpunct}{\mcitedefaultseppunct}\relax
\EndOfBibitem
\bibitem[Boogert \latin{et~al.}(2015)Boogert, Gerakines, and
  Whittet]{BoogertARAA15}
Boogert,~A.~C.~A.; Gerakines,~P.~A.; Whittet,~D. C.~B. {Observations of the icy
  universe.} \emph{Annual Rev. Astron. Astrophys.} \textbf{2015}, \emph{53},
  541--581\relax
\mciteBstWouldAddEndPuncttrue
\mciteSetBstMidEndSepPunct{\mcitedefaultmidpunct}
{\mcitedefaultendpunct}{\mcitedefaultseppunct}\relax
\EndOfBibitem
\bibitem[{M{\"u}ller} \latin{et~al.}(2001){M{\"u}ller}, {Thorwirth}, {Roth},
  and {Winnewisser}]{MullerAA01}
{M{\"u}ller},~H.~S.~P.; {Thorwirth},~S.; {Roth},~D.~A.; {Winnewisser},~G. {The
  Cologne Database for Molecular Spectroscopy, CDMS}. \emph{Astronomy and
  Astrophysics} \textbf{2001}, \emph{370}, L49--L52\relax
\mciteBstWouldAddEndPuncttrue
\mciteSetBstMidEndSepPunct{\mcitedefaultmidpunct}
{\mcitedefaultendpunct}{\mcitedefaultseppunct}\relax
\EndOfBibitem
\bibitem[{Theul{\'e}} \latin{et~al.}(2013){Theul{\'e}}, {Duvernay}, {Danger},
  {Borget}, {Bossa}, {Vinogradoff}, {Mispelaer}, and {Chiavassa}]{TheuleASR13}
{Theul{\'e}},~P.; {Duvernay},~F.; {Danger},~G.; {Borget},~F.; {Bossa},~J.~B.;
  {Vinogradoff},~V.; {Mispelaer},~F.; {Chiavassa},~T. {Thermal reactions in
  interstellar ice: A step towards molecular complexity in the interstellar
  medium}. \emph{Advances in Space Research} \textbf{2013}, \emph{52},
  1567--1579\relax
\mciteBstWouldAddEndPuncttrue
\mciteSetBstMidEndSepPunct{\mcitedefaultmidpunct}
{\mcitedefaultendpunct}{\mcitedefaultseppunct}\relax
\EndOfBibitem
\bibitem[Linnartz \latin{et~al.}(2015)Linnartz, Ioppolo, and
  Fedoseev]{LinnartzIRPC15}
Linnartz,~H.; Ioppolo,~S.; Fedoseev,~G. Atom addition reactions in interstellar
  ice analogues. \emph{International Reviews in Physical Chemistry}
  \textbf{2015}, \emph{34}, 205--223\relax
\mciteBstWouldAddEndPuncttrue
\mciteSetBstMidEndSepPunct{\mcitedefaultmidpunct}
{\mcitedefaultendpunct}{\mcitedefaultseppunct}\relax
\EndOfBibitem
\bibitem[Oberg({2016})]{ObergCR16}
Oberg,~K.~I. {Photochemistry and Astrochemistry: Photochemical Pathways to
  Interstellar Complex Organic Molecules}. \emph{{Chemical Reviews}}
  \textbf{{2016}}, \emph{{116}}, {9631--9663}\relax
\mciteBstWouldAddEndPuncttrue
\mciteSetBstMidEndSepPunct{\mcitedefaultmidpunct}
{\mcitedefaultendpunct}{\mcitedefaultseppunct}\relax
\EndOfBibitem
\bibitem[{Potapov} \latin{et~al.}(2017){Potapov}, {J{\"a}ger}, {Henning},
  {Jonusas}, and {Krim}]{PotapovApJ17}
{Potapov},~A.; {J{\"a}ger},~C.; {Henning},~T.; {Jonusas},~M.; {Krim},~L. {The
  Formation of Formaldehyde on Interstellar Carbonaceous Grain Analogs by O/H
  Atom Addition}. \emph{The Astrophysical Journal} \textbf{2017}, \emph{846},
  131\relax
\mciteBstWouldAddEndPuncttrue
\mciteSetBstMidEndSepPunct{\mcitedefaultmidpunct}
{\mcitedefaultendpunct}{\mcitedefaultseppunct}\relax
\EndOfBibitem
\bibitem[Fraser \latin{et~al.}({2002})Fraser, Collings, and
  McCoustra]{FraserRSI02}
Fraser,~H.; Collings,~M.; McCoustra,~M. {Laboratory surface astrophysics
  experiment}. \emph{{Review of Scientific Instruments}} \textbf{{2002}},
  \emph{{73}}, {2161--2170}\relax
\mciteBstWouldAddEndPuncttrue
\mciteSetBstMidEndSepPunct{\mcitedefaultmidpunct}
{\mcitedefaultendpunct}{\mcitedefaultseppunct}\relax
\EndOfBibitem
\bibitem[Abplanalp \latin{et~al.}({2016})Abplanalp, Foerstel, and
  Kaiser]{AbplanaplCPL16}
Abplanalp,~M.~J.; Foerstel,~M.; Kaiser,~R.~I. {Exploiting single photon vacuum
  ultraviolet photoionization to unravel the synthesis of complex organic
  molecules in interstellar ices}. \emph{{Chemical Physics Letters}}
  \textbf{{2016}}, \emph{{644}}, {79--98}\relax
\mciteBstWouldAddEndPuncttrue
\mciteSetBstMidEndSepPunct{\mcitedefaultmidpunct}
{\mcitedefaultendpunct}{\mcitedefaultseppunct}\relax
\EndOfBibitem
\bibitem[{Coustenis} and {Hirtzig}(2009){Coustenis}, and
  {Hirtzig}]{CoustenisRAA09}
{Coustenis},~A.; {Hirtzig},~M. {Cassini-Huygens results on Titan's surface}.
  \emph{Research in Astronomy and Astrophysics} \textbf{2009}, \emph{9},
  249--268\relax
\mciteBstWouldAddEndPuncttrue
\mciteSetBstMidEndSepPunct{\mcitedefaultmidpunct}
{\mcitedefaultendpunct}{\mcitedefaultseppunct}\relax
\EndOfBibitem
\bibitem[{Kr{\"u}ger} \latin{et~al.}(2015){Kr{\"u}ger}, {Stephan}, {Engrand},
  {Briois}, {Siljestr{\"o}m}, {Merouane}, {Baklouti}, {Fischer}, {Fray}, and
  {Hornung}]{KrugerPSS15}
{Kr{\"u}ger},~H.; {Stephan},~T.; {Engrand},~C.; {Briois},~C.;
  {Siljestr{\"o}m},~S.; {Merouane},~S.; {Baklouti},~D.; {Fischer},~H.;
  {Fray},~N.; {Hornung},~K. {COSIMA-Rosetta calibration for in situ
  characterization of 67P/Churyumov-Gerasimenko cometary inorganic compounds}.
  \emph{Planetary Space Science} \textbf{2015}, \emph{117}, 35--44\relax
\mciteBstWouldAddEndPuncttrue
\mciteSetBstMidEndSepPunct{\mcitedefaultmidpunct}
{\mcitedefaultendpunct}{\mcitedefaultseppunct}\relax
\EndOfBibitem
\bibitem[{Altwegg} \latin{et~al.}(2017){Altwegg}, {Balsiger}, {Berthelier},
  {Bieler}, {Calmonte}, {Fuselier}, {Goesmann}, {Gasc}, {Gombosi}, and {Le
  Roy}]{AltweggMNRAS17}
{Altwegg},~K.; {Balsiger},~H.; {Berthelier},~J.~J.; {Bieler},~A.;
  {Calmonte},~U.; {Fuselier},~S.~A.; {Goesmann},~F.; {Gasc},~S.;
  {Gombosi},~T.~I.; {Le Roy},~L. {Organics in comet 67P - a first comparative
  analysis of mass spectra from ROSINA-DFMS, COSAC and Ptolemy}. \emph{Monthly
  Notices of the Royal Astronomical Society} \textbf{2017}, \emph{469},
  S130--S141\relax
\mciteBstWouldAddEndPuncttrue
\mciteSetBstMidEndSepPunct{\mcitedefaultmidpunct}
{\mcitedefaultendpunct}{\mcitedefaultseppunct}\relax
\EndOfBibitem
\bibitem[{Altwegg} \latin{et~al.}(2016){Altwegg}, {Balsiger}, {Bar-Nun},
  {Berthelier}, {Bieler}, {Bochsler}, {Briois}, {Calmonte}, {Combi}, and
  {Cottin}]{AltweggSA16}
{Altwegg},~K.; {Balsiger},~H.; {Bar-Nun},~A.; {Berthelier},~J.~J.;
  {Bieler},~A.; {Bochsler},~P.; {Briois},~C.; {Calmonte},~U.; {Combi},~M.~R.;
  {Cottin},~H. {Prebiotic chemicals--amino acid and phosphorus--in the coma of
  comet 67P/Churyumov-Gerasimenko}. \emph{Science Advances} \textbf{2016},
  \emph{2}, e1600285--e1600285\relax
\mciteBstWouldAddEndPuncttrue
\mciteSetBstMidEndSepPunct{\mcitedefaultmidpunct}
{\mcitedefaultendpunct}{\mcitedefaultseppunct}\relax
\EndOfBibitem
\bibitem[{Yocum} \latin{et~al.}(2019){Yocum}, {Smith}, {Todd}, {Mora},
  {Gerakines}, {Milam}, and {Widicus Weaver}]{YocumJCPA19}
{Yocum},~K.; {Smith},~H.~H.; {Todd},~E.; {Mora},~L.; {Gerakines},~P.~A.;
  {Milam},~S.~N.; {Widicus Weaver},~S.~L. {Millimeter/Submillimeter
  Spectroscopic Detection of Desorbed Ices: A New Technique in Laboratory
  Astrochemistry}. \emph{J. Phys. Chem. A} \textbf{2019}, \emph{123},
  8702--8708\relax
\mciteBstWouldAddEndPuncttrue
\mciteSetBstMidEndSepPunct{\mcitedefaultmidpunct}
{\mcitedefaultendpunct}{\mcitedefaultseppunct}\relax
\EndOfBibitem
\bibitem[Baratta \latin{et~al.}({2002})Baratta, Leto, and Palumbo]{BarattaAA02}
Baratta,~G.; Leto,~G.; Palumbo,~M. {A comparison of ion irradiation and UV
  photolysis of CH$_4$ and CH$_3$OH}. \emph{{Astronomy \& Astrophysics}}
  \textbf{{2002}}, \emph{{384}}, {343--349}\relax
\mciteBstWouldAddEndPuncttrue
\mciteSetBstMidEndSepPunct{\mcitedefaultmidpunct}
{\mcitedefaultendpunct}{\mcitedefaultseppunct}\relax
\EndOfBibitem
\bibitem[Bertin \latin{et~al.}({2009})Bertin, Martin, Duvernay, Theule, Bossa,
  Borget, Illenberger, Lafosse, Chiavassa, and Azria]{BertinPCCP09}
Bertin,~M.; Martin,~I.; Duvernay,~F.; Theule,~P.; Bossa,~J.~B.; Borget,~F.;
  Illenberger,~E.; Lafosse,~A.; Chiavassa,~T.; Azria,~R. {Chemistry induced by
  low-energy electrons in condensed multilayers of ammonia and carbon dioxide}.
  \emph{{Physical Chemistry Chemical Physics}} \textbf{{2009}}, \emph{{11}},
  {1838--1845}\relax
\mciteBstWouldAddEndPuncttrue
\mciteSetBstMidEndSepPunct{\mcitedefaultmidpunct}
{\mcitedefaultendpunct}{\mcitedefaultseppunct}\relax
\EndOfBibitem
\bibitem[{Bossa} \latin{et~al.}(2009){Bossa}, {Duvernay}, {Theul{\'e}},
  {Borget}, {D'Hendecourt}, and {Chiavassa}]{BossaAA09}
{Bossa},~J.-B.; {Duvernay},~F.; {Theul{\'e}},~P.; {Borget},~F.;
  {D'Hendecourt},~L.; {Chiavassa},~T. {Methylammonium methylcarbamate thermal
  formation in interstellar ice analogs: a glycine salt precursor in
  protostellar environments}. \emph{Astronomy and Astrophysics} \textbf{2009},
  \emph{506}, 601--608\relax
\mciteBstWouldAddEndPuncttrue
\mciteSetBstMidEndSepPunct{\mcitedefaultmidpunct}
{\mcitedefaultendpunct}{\mcitedefaultseppunct}\relax
\EndOfBibitem
\bibitem[{Bossa} \latin{et~al.}(2008){Bossa}, {Theul{\'e}}, {Duvernay},
  {Borget}, and {Chiavassa}]{BossaAA08}
{Bossa},~J.~B.; {Theul{\'e}},~P.; {Duvernay},~F.; {Borget},~F.; {Chiavassa},~T.
  {Carbamic acid and carbamate formation in NH$_3$:CO$_2$ ices - UV irradiation
  versus thermal processes}. \emph{Astronomy and Astrophysics} \textbf{2008},
  \emph{492}, 719--724\relax
\mciteBstWouldAddEndPuncttrue
\mciteSetBstMidEndSepPunct{\mcitedefaultmidpunct}
{\mcitedefaultendpunct}{\mcitedefaultseppunct}\relax
\EndOfBibitem
\bibitem[{Potapov} \latin{et~al.}(2019){Potapov}, {Theul{\'e}}, {J{\"a}ger},
  and {Henning}]{PotapovApJ19}
{Potapov},~A.; {Theul{\'e}},~P.; {J{\"a}ger},~C.; {Henning},~T. {Evidence of
  Surface Catalytic Effect on Cosmic Dust Grain Analogs: The Ammonia and Carbon
  Dioxide Surface Reaction}. \emph{The Astrophysical Journal Letters}
  \textbf{2019}, \emph{878}, L20\relax
\mciteBstWouldAddEndPuncttrue
\mciteSetBstMidEndSepPunct{\mcitedefaultmidpunct}
{\mcitedefaultendpunct}{\mcitedefaultseppunct}\relax
\EndOfBibitem
\bibitem[Hermanns \latin{et~al.}(2019)Hermanns, Wehres, Lewen, Müller, and
  Schlemmer]{HermannsJMS19}
Hermanns,~M.; Wehres,~N.; Lewen,~F.; Müller,~H.; Schlemmer,~S. Rotational
  spectroscopy of the two higher energy conformers of 2-cyanobutane.
  \emph{Journal of Molecular Spectroscopy} \textbf{2019}, \emph{358}, 25 --
  36\relax
\mciteBstWouldAddEndPuncttrue
\mciteSetBstMidEndSepPunct{\mcitedefaultmidpunct}
{\mcitedefaultendpunct}{\mcitedefaultseppunct}\relax
\EndOfBibitem
\bibitem[Brown \latin{et~al.}(2008)Brown, Dian, Douglass, Geyer, Shipman, and
  Pate]{BrownRSI08}
Brown,~G.~G.; Dian,~B.~C.; Douglass,~K.~O.; Geyer,~S.~M.; Shipman,~S.~T.;
  Pate,~B.~H. Broadband Fourier transform microwave spectrometer based on
  chirped pulse excitation. \emph{Review of Scientific Instruments}
  \textbf{2008}, \emph{79}, 053103\relax
\mciteBstWouldAddEndPuncttrue
\mciteSetBstMidEndSepPunct{\mcitedefaultmidpunct}
{\mcitedefaultendpunct}{\mcitedefaultseppunct}\relax
\EndOfBibitem
\bibitem[Balle and Flygare(1981)Balle, and Flygare]{BalleRSI81}
Balle,~T.; Flygare,~W. Fabry-Perot cavity pulsed Fourier-transform microwave
  spectrometer with a pulsed nozzle particle source. \emph{Review of Scientific
  Instruments} \textbf{1981}, \emph{52}, 33--45\relax
\mciteBstWouldAddEndPuncttrue
\mciteSetBstMidEndSepPunct{\mcitedefaultmidpunct}
{\mcitedefaultendpunct}{\mcitedefaultseppunct}\relax
\EndOfBibitem
\bibitem[Park \latin{et~al.}(2011)Park, Steeves, Kuyanov-Prozument, Neill, and
  Field]{ParkJCP11}
Park,~G.~B.; Steeves,~A.~H.; Kuyanov-Prozument,~K.; Neill,~J.~L.; Field,~R.~W.
  Design and evaluation of a pulsed-jet chirped-pulse millimeter-wave
  spectrometer for the 70-102 GHz region. \emph{Journal of Chemical Physics}
  \textbf{2011}, \emph{135}, 024202\relax
\mciteBstWouldAddEndPuncttrue
\mciteSetBstMidEndSepPunct{\mcitedefaultmidpunct}
{\mcitedefaultendpunct}{\mcitedefaultseppunct}\relax
\EndOfBibitem
\bibitem[Jahn \latin{et~al.}(2012)Jahn, Dewald, Wachsmuth, Grabow, and
  Mehrotra]{JahnJMS12}
Jahn,~M.~K.; Dewald,~D.~A.; Wachsmuth,~D.; Grabow,~J.-U.; Mehrotra,~S.~C.
  {Rapid capture of large amplitude motions in 2,6-difluorophenol:
  High-resolution fast-passage FT-MW technique}. \emph{Journal of Molecular
  Spectroscopy} \textbf{2012}, \emph{280}, 54 -- 60\relax
\mciteBstWouldAddEndPuncttrue
\mciteSetBstMidEndSepPunct{\mcitedefaultmidpunct}
{\mcitedefaultendpunct}{\mcitedefaultseppunct}\relax
\EndOfBibitem
\bibitem[Steber \latin{et~al.}(2012)Steber, Harris, Neill, and
  Pate]{SteberJMS12}
Steber,~A.~L.; Harris,~B.~J.; Neill,~J.~L.; Pate,~B.~H. {An arbitrary waveform
  generator based chirped pulse Fourier transform spectrometer operating from
  260 to 295 GHz}. \emph{{Journal of Molecular Spectroscopy}} \textbf{2012},
  \emph{{280}}, 3 -- 10, Broadband Rotational Spectroscopy\relax
\mciteBstWouldAddEndPuncttrue
\mciteSetBstMidEndSepPunct{\mcitedefaultmidpunct}
{\mcitedefaultendpunct}{\mcitedefaultseppunct}\relax
\EndOfBibitem
\bibitem[Oldham \latin{et~al.}({2014})Oldham, Abeysekera, Joalland, Zack,
  Prozument, Sims, Park, Field, and Suits]{OldhamJCP14}
Oldham,~J.~M.; Abeysekera,~C.; Joalland,~B.; Zack,~L.~N.; Prozument,~K.;
  Sims,~I.~R.; Park,~G.~B.; Field,~R.~W.; Suits,~A.~G. {A chirped-pulse
  Fourier-transform microwave/pulsed uniform flow spectrometer. I. The
  low-temperature flow system}. \emph{{Journal of Chemical Physics}}
  \textbf{{2014}}, \emph{{141}}, 154202\relax
\mciteBstWouldAddEndPuncttrue
\mciteSetBstMidEndSepPunct{\mcitedefaultmidpunct}
{\mcitedefaultendpunct}{\mcitedefaultseppunct}\relax
\EndOfBibitem
\bibitem[Prozument \latin{et~al.}(2014)Prozument, Park, Shaver, Vasiliou,
  Oldham, David, Muenter, Stanton, Suits, Ellison, and Field]{ProzumentPCCP14}
Prozument,~K.; Park,~G.~B.; Shaver,~R.~G.; Vasiliou,~A.~K.; Oldham,~J.~M.;
  David,~D.~E.; Muenter,~J.~S.; Stanton,~J.~F.; Suits,~A.~G.; Ellison,~G.~B.;
  Field,~R.~W. Chirped-pulse millimeter-wave spectroscopy for dynamics and
  kinetics studies of pyrolysis reactions. \emph{Physical Chemistry Chemical
  Physics} \textbf{2014}, \emph{16}, 15739--15751\relax
\mciteBstWouldAddEndPuncttrue
\mciteSetBstMidEndSepPunct{\mcitedefaultmidpunct}
{\mcitedefaultendpunct}{\mcitedefaultseppunct}\relax
\EndOfBibitem
\bibitem[Collings \latin{et~al.}({2004})Collings, Anderson, Chen, Dever, Viti,
  Williams, and McCoustra]{CollingsMNRAS04}
Collings,~M.; Anderson,~M.; Chen,~R.; Dever,~J.; Viti,~S.; Williams,~D.;
  McCoustra,~M. {A laboratory survey of the thermal desorption of
  astrophysically relevant molecules}. \emph{{Monthly Notices of the Royal
  Astronomical Society}} \textbf{{2004}}, \emph{{354}}, {1133--1140}\relax
\mciteBstWouldAddEndPuncttrue
\mciteSetBstMidEndSepPunct{\mcitedefaultmidpunct}
{\mcitedefaultendpunct}{\mcitedefaultseppunct}\relax
\EndOfBibitem
\bibitem[{Sandford} and {Allamandola}(1993){Sandford}, and
  {Allamandola}]{SandfordApJ93}
{Sandford},~S.~A.; {Allamandola},~L.~J. {Condensation and vaporization studies
  of CH$_3$OH and NH$_3$ ices: Major implications for astrochemistry}.
  \emph{The Astrophysical Journal} \textbf{1993}, \emph{417}, 815--825\relax
\mciteBstWouldAddEndPuncttrue
\mciteSetBstMidEndSepPunct{\mcitedefaultmidpunct}
{\mcitedefaultendpunct}{\mcitedefaultseppunct}\relax
\EndOfBibitem
\bibitem[{Fraser} \latin{et~al.}(2001){Fraser}, {Collings}, {McCoustra}, and
  {Williams}]{FraserMNRAS01}
{Fraser},~H.~J.; {Collings},~M.~P.; {McCoustra},~M. R.~S.; {Williams},~D.~A.
  {Thermal desorption of water ice in the interstellar medium}. \emph{Mon. Not.
  Roy. Ast. Soc.} \textbf{2001}, \emph{327}, 1165--1172\relax
\mciteBstWouldAddEndPuncttrue
\mciteSetBstMidEndSepPunct{\mcitedefaultmidpunct}
{\mcitedefaultendpunct}{\mcitedefaultseppunct}\relax
\EndOfBibitem
\bibitem[{Acharyya} \latin{et~al.}(2007){Acharyya}, {Fuchs}, {Fraser}, {van
  Dishoeck}, and {Linnartz}]{AcharyyaAA07}
{Acharyya},~K.; {Fuchs},~G.~W.; {Fraser},~H.~J.; {van Dishoeck},~E.~F.;
  {Linnartz},~H. {Desorption of CO and O$_{2}$ interstellar ice analogs}.
  \emph{Astronomy \& Astrophysics} \textbf{2007}, \emph{466}, 1005--1012\relax
\mciteBstWouldAddEndPuncttrue
\mciteSetBstMidEndSepPunct{\mcitedefaultmidpunct}
{\mcitedefaultendpunct}{\mcitedefaultseppunct}\relax
\EndOfBibitem
\bibitem[{Noble} \latin{et~al.}(2012){Noble}, {Theule}, {Mispelaer},
  {Duvernay}, {Danger}, {Congiu}, {Dulieu}, and {Chiavassa}]{NobleAA12}
{Noble},~J.~A.; {Theule},~P.; {Mispelaer},~F.; {Duvernay},~F.; {Danger},~G.;
  {Congiu},~E.; {Dulieu},~F.; {Chiavassa},~T. {The desorption of H$_{2}$CO from
  interstellar grains analogues}. \emph{Astronomy and Astrophysics}
  \textbf{2012}, \emph{543}, A5\relax
\mciteBstWouldAddEndPuncttrue
\mciteSetBstMidEndSepPunct{\mcitedefaultmidpunct}
{\mcitedefaultendpunct}{\mcitedefaultseppunct}\relax
\EndOfBibitem
\bibitem[Readhead(1962)]{RedheadV62}
Readhead,~J. Thermal desorption of gases. \emph{Vacuum} \textbf{1962},
  \emph{12}, 203\relax
\mciteBstWouldAddEndPuncttrue
\mciteSetBstMidEndSepPunct{\mcitedefaultmidpunct}
{\mcitedefaultendpunct}{\mcitedefaultseppunct}\relax
\EndOfBibitem
\bibitem[Abeysekera \latin{et~al.}({2014})Abeysekera, Zack, Park, Joalland,
  Oldham, Prozument, Ariyasingha, Sims, Field, and Suits]{AbeysekeraJCP14}
Abeysekera,~C.; Zack,~L.~N.; Park,~G.~B.; Joalland,~B.; Oldham,~J.~M.;
  Prozument,~K.; Ariyasingha,~N.~M.; Sims,~I.~R.; Field,~R.~W.; Suits,~A.~G. {A
  chirped-pulse Fourier-transform microwave/pulsed uniform flow spectrometer.
  II. Performance and applications for reaction dynamics}. \emph{{Journal of
  Chemical Physics}} \textbf{{2014}}, \emph{{141}}, {214203}\relax
\mciteBstWouldAddEndPuncttrue
\mciteSetBstMidEndSepPunct{\mcitedefaultmidpunct}
{\mcitedefaultendpunct}{\mcitedefaultseppunct}\relax
\EndOfBibitem
\bibitem[{Park, G. Barratt and Field, Robert W.}({2015})]{ParkJMS15}
{Park, G. Barratt and Field, Robert W.}, {Edge effects in chirped-pulse Fourier
  transform microwave spectra}. \emph{{Journal of Molecular Spectroscopy}}
  \textbf{{2015}}, \emph{{312}}, {54--57}\relax
\mciteBstWouldAddEndPuncttrue
\mciteSetBstMidEndSepPunct{\mcitedefaultmidpunct}
{\mcitedefaultendpunct}{\mcitedefaultseppunct}\relax
\EndOfBibitem
\end{mcitethebibliography}

\end{document}